\documentclass[aps,10pt,prb,twocolumn,showpacs,superscriptaddress,citeautoscript]{revtex4-1}

\usepackage[utf8]{inputenc}
\usepackage[colorlinks]{hyperref}	
\usepackage{amsfonts}
\usepackage{epsfig,xcolor}
\usepackage{bm}
\usepackage{amsmath}
\usepackage{units}

\usepackage{color} % For comments

\newcommand{\vc}[1]{\bm{#1}}

\renewcommand{\k}{{\vc k}}

\newcommand{\K}{{\vc K}}
\newcommand{\vF}{{v_{\rm F}}}
\newcommand{\kB}{{k_{\rm B}}}

\newcommand\ie{\textit{i.e.}}

\DeclareMathOperator{\Tr}{Tr}
\newcommand{\DepFisMat}{GISC, Departamento de F\'{\i}sica de Materiales,
Universidad Complutense, E-28040 Madrid, Spain}

\begin{document}

\title{
Strong spin-dependent negative differential resistance in composite graphene superlattices
}

\author{J. Mun\'{a}rriz}
\affiliation{\DepFisMat}

\author{C. Gaul}
\affiliation{\DepFisMat}
\affiliation{CEI Campus Moncloa, UCM-UPM, Madrid, Spain}
\affiliation{Max Planck Institute for the Physics of Complex Systems, 01187
Dresden, Germany}

\author{A. V. Malyshev}
\affiliation{\DepFisMat}
\affiliation{Ioffe Physical-Technical Institute, 26 Politechnicheskaya str.,
194021 St-Petersburg, Russia}

\author{P. A. Orellana}
\affiliation{Departamento de F\'{\i}sica, Universidad T\'{e}cnica Federico
Santa Mar\'{\i}a, Casilla 110 V, Valpara\'{\i}so, Chile}

\author{C. A. M\"{u}ller}
\affiliation{Centre for Quantum Technologies, National University of Singapore,
Singapore 117543, Singapore}
\affiliation{Department of Physics, University of Konstanz, Germany}

\author{F. Dom\'{\i}nguez-Adame}
\affiliation{\DepFisMat}

\begin{abstract}

We find clear signatures of spin-dependent negative differential resistance in
compound systems comprising a graphene nanoribbon and a set of ferromagnetic
insulator strips deposited on top of it.  The periodic array of ferromagnetic
strips induces a proximity exchange splitting of the electronic states in
graphene, resulting in the appearance of a superlattice with a spin-dependent
energy spectrum.  The electric current through the device can be highly
polarized and both the current and its polarization manifest non-monotonic
dependence on the bias voltage.  The device operates therefore as an Esaki
spin diode, which opens possibilities to design new spintronic circuits.

\end{abstract}


\maketitle

\section{Introduction}

Since the pioneering work by Esaki~\cite{Esaki1958}, quantum tunneling and
negative differential resistance (NDR) have been the underlying principle of
operation of various quantum devices~\cite{Esaki1974,Sze2006,Malyshev2007}.  NDR
is often related to the resonant tunneling of carriers; when the
chemical potential of a lead approaches one of the resonant levels of a
device, the current $I$ increases. However, the resonant level position can depend
on the applied voltage $V$, which can finally drive the system out of
resonance.  Then, the current decreases dramatically with a further
increase of the voltage. The resulting $I$-$V$ characteristics are typically
N-shaped and include a region with NDR. Such a conductance anomaly can, for example, be
observed in semiconductor heterostructures~\cite{Esaki1974},  semiconductor
superlattices,\cite{Tsu1973}  conductor/superconductor
junctions,\cite{Kummel1990}  carbon nanotubes,\cite{Leonard2000}  molecular
systems~\cite{Malyshev2007} and at the atomic scale~\cite{Bedrossian1989}.

Due to its remarkable charge transport properties~\cite{Novoselov2004} and long
spin-coherence length~\cite{Kane2005,Tombros2007,Han2012,Zomer2012,Lundeberg2013}, graphene is a very promising material for
spintronics~\cite{Rozhkov2011,Munarriz2012}.  Graphene nanoribbons (GNR) with
tailored edges (zigzag or armchair) provide means to generate and manipulate
spin-polarized electrons.\cite{Wakabayashi2009}  In this regard, signatures of
NDR for spin-down electrons in Be-doped zigzag GNRs have already been found by
Wu \emph{et al.}\cite{Wu2012b}, where spin-polarized edge states play an
important role.

Here, we consider a spin-dependent superlattice realized by ferromagnetic 
insulator strips\cite{Haugen2008} deposited on top of an armchair GNR. 
Similar proposals on (ferromagnetic) superlattices of graphene have been
presented recently. Yu \emph{et al.}\cite{Yu2012a} have studied a superlattice
realized by stubs in the shape of a zigzag GNR with a ferromagnetic insulator on
top of the whole system. They found strongly spin-dependent minibands and
minigaps, but they did not study the effect of a bias voltage, nor have they
found NDR. Niu \emph{et al.}\cite{Niu2008} and Faizabadi \emph{et
al.}\cite{Faizabadi2012} have investigated a superlattice made of gated
ferromagnetic strips on top of graphene.  However, the finite width of the GNR
and the quantization of the transverse momentum was not taken into account.
Instead, they took the incident angle as a free parameter.  They found that spin
polarization of tunneling conductance and magnetoresistance exhibit oscillatory
behavior as a function of the gate voltage, but they did not consider the bias
voltage either. Finally, Ferreira \emph{et al.}\cite{Ferreira2011} studied an
armchair GNR under a spin-independent superlattice and a bias field, which leads
to a spin-independent NDR effect.

In this paper we propose a graphene-based device whose $I$-$V$ characteristics
show spin-dependent NDR with high peak-to-valley ratios, which could be an
important building block for future spintronic devices. The structure of the
paper is as follows. In Sec.~\ref{secModel} we present the setup of a gapped
armchair GNR with several strips of a ferromagnetic insulator on top of it,
which creates a spin-dependent superlattice. 
We compute the stationary
wave function across the sample and the transmission coefficient for a given
spin, energy, and bias voltage. The resulting current-voltage characteristics of
the device, comprising a spin-filtering effect and a strong spin-dependent NDR,
are discussed in the subsequent Sec.~\ref{secResults}, while
Sec.~\ref{conclusions} concludes the paper and provides an outlook on possible
further developments.

\section{%Model 
Setup and formalism}\label{secModel}

\begin{figure}[bt]
\begin{center}
\includegraphics[width=0.90\columnwidth,clip]{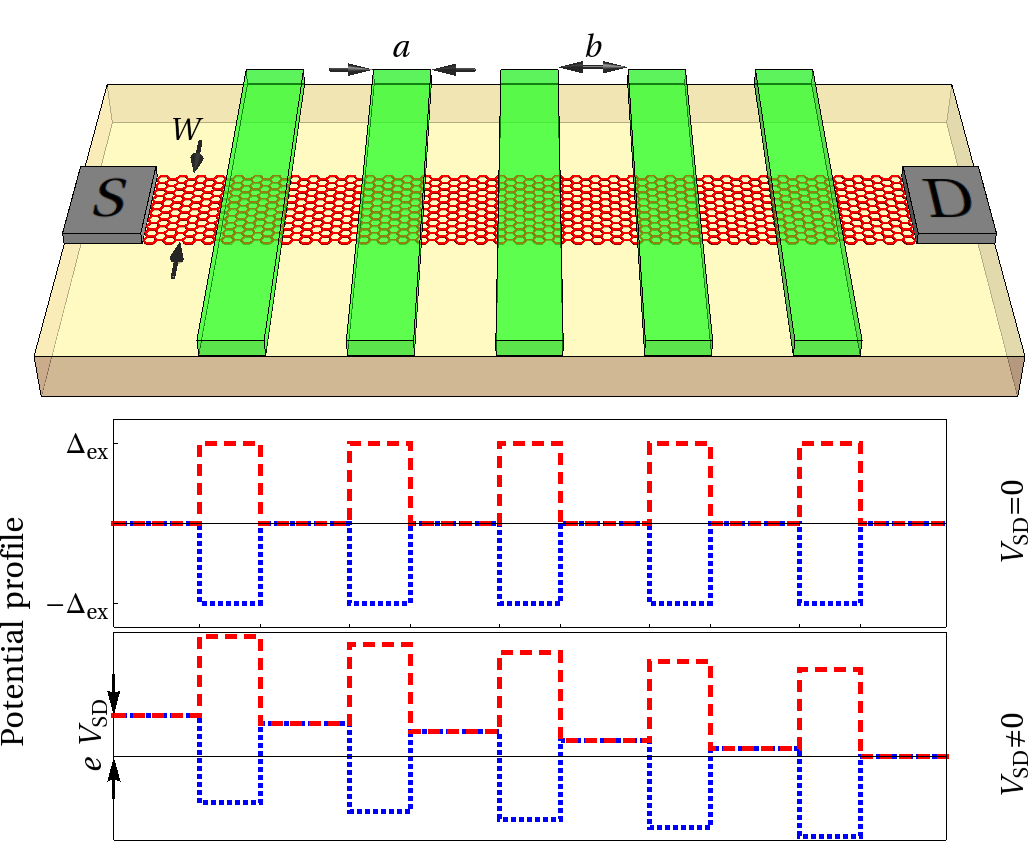}
\end{center}
\caption{(Color online) % Setup of the device.
The upper panel shows the GNR connected to source~(S) and drain~(D) leads, with
$N=5$ perpendicular strips of a ferromagnetic insulator (green bars) on top of
it. The model potential profiles for spin-up (dashed
red lines) and spin-down (dotted blue lines) electrons in the unbiased and biased device are shown in the middle
and lower panels respectively.
}
\label{figScheme}
\end{figure}

The proposed system is composed of a rectangular GNR of width $W\simeq
\unit[9.8]{nm}$, connected to source and drain leads, and $N=5$ rectangular
strips of a ferromagnetic insulator arranged periodically on top of the GNR (see
the upper panel of Fig.~\ref{figScheme}). As we discuss later, this number of
ferromagnetic strips is enough to reveal clear signatures of spin-dependent
NDR. The width of the strips is $a=\unit[23.9]{nm}$ and the spacing between them
is $b=\unit[55.8]{nm}$. It is known that both the width and the edge type of a
GNR strongly affect its electronic properties. Experimental
evidences~\cite{Han2007} and \emph{ab-initio} calculations~\cite{Son2006} show
that the spectrum of a GNR with armchair edges has a gap, which is inversely
proportional to the width $W$ and depends on the remainder $(2W/a_0 \mod 3)$,
where $a_0 = \unit[0.246]{nm}$ is the lattice constant, \ie, the width of the graphene lattice hexagon.
Contrary to GNRs with zigzag edges, the dispersion relation of the armchair GNR
is centered around $k = 0$. This  is advantageous for tunneling structures
because the resonant levels are expected to be broader and less affected by
disorder~\cite{Munarriz2011}. We therefore restrict ourselves to the armchair GNRs.

EuO can be used as the ferromagnetic insulator for the superlattice; this
material has been studied in conjunction with graphene both
experimentally~\cite{Forster2011,Swartz2012} and
theoretically~\cite{Haugen2008}. The  proximity
exchange interaction between magnetic ions in the strips and charge carriers in
the GNR can be described as an effective Zeeman splitting
$\pm\Delta_\mathrm{ex}$ of the spin sublevels~\cite{Haugen2008}.  There is still
no consensus on the magnitude of the exchange splitting amplitude
$\Delta_\mathrm{ex}$ in graphene.  We use $\Delta_\mathrm{ex}=\unit[5]{meV}$,
which lies in the range of values known from the literature
($\unit[3-10]{meV}$)~\cite{Haugen2008,Zou2009,Gu2009}.  We have checked that our
results do not change qualitatively if we use a different value of
$\Delta_\mathrm{ex}$ within the known range.

Because the proximity exchange interaction has the characteristic length scale
of one atomic layer, the splitting is induced only in the regions of the GNR
which are just below the ferromagnetic strips. Therefore, for the chosen system
geometry, a spin-up (spin-down) electron propagating along the sample will be
subjected to a potential comprising a periodic set of rectangular barriers
(wells), as plotted in the middle panel of Fig.~\ref{figScheme}.  In other
words, the array of the ferromagnetic strips creates a spin-dependent
superlattice. We note that similar systems manifesting NDR have been studied in
Ref.~\cite{Ferreira2011}, but the superlattice potential was supposed to be
induced by electrostatic gates, so all characteristics were spin independent.

\subsection{Tight-binding method}\label{secTB}

A simple tight-binding Hamiltonian of a single electron in the $p_z$ orbitals
of
graphene is widely used to model GNRs. For low energy excitations, \ie, energies
close to the Dirac point, hopping %the interaction 
can be restricted to the nearest
neighbors.  Then, the Hamiltonian can be written as
\begin{equation}
\mathcal{H}= 
- t\sum_{\langle i,j\rangle} |i\rangle\langle j|
+\sum_{i}\epsilon_i |i \rangle\langle i|
+ \sigma\,\Delta_\mathrm{ex}\sum_{i\in {\mathcal L}}|i \rangle\langle i| \ . 
\end{equation}
Here $|i \rangle$ is the ket vector of the atomic orbital of the $i$th carbon
atom, $t = \unit[2.8]{eV}$ is the hopping between neighboring atoms, the full
set of which is denoted as $\langle i,j\rangle$.  The on-site energy is the
sum %perposition 
of the following two terms: the bias-induced electrostatic
potential $\epsilon_i$  at the position of the $i$th atom (see
Sec.~\ref{secBias}) and the spin-dependent exchange-interaction %shift
$\Delta_\mathrm{ex}$ due to the ferromagnetic strips, with $\sigma=\pm 1$ for
spin-up and spin-down electrons. The exchange-interaction is induced only at
the atoms that are in direct contact with the ferromagnetic strips (the full set of
them is labeled as $\cal L$ in the above equation). The on-site energy is sketched, for zero and finite bias, in the middle and lower panels of Fig.~\ref{figScheme}, respectively. 

The wave function in the GNR can be obtained using the quantum transmission
boundary method \cite{Lent1990,Ting1992}. This is accomplished by assuming
semi-infinite leads, whose modes are calculated using an effective
transfer-matrix approach \cite{Schelter2010}. Then, both the ingoing and
outgoing wave functions are computed as linear combinations of propagating plane
waves at a given energy, and the corresponding amplitudes determine the
spin-dependent transmission probabilities $T_\pm$.

\subsection{Dirac theory}

\begin{figure} %[ht]
\includegraphics[width=\linewidth]{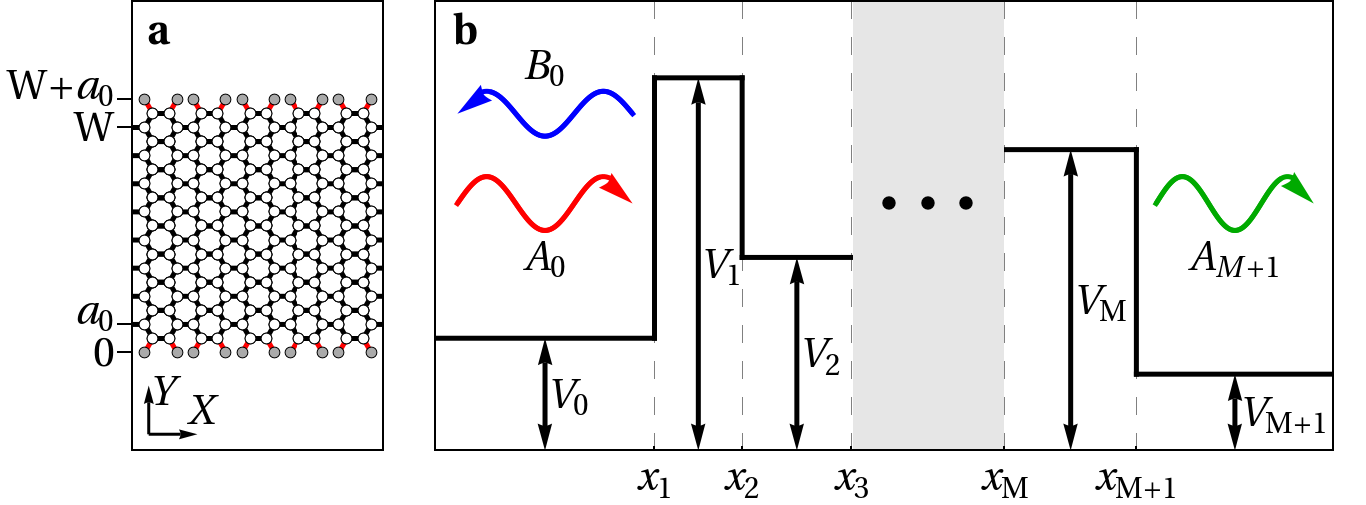}
\caption{(Color online)
(a)~Scheme of a GNR of width $W=8 a_0$. The boundary  conditions for wave
functions can be obtained by adding two rows of atoms (plotted  in gray) at
$y=0$ and $W+a_0$ and setting the wavefunction to $0$ in those points.
(b)~Transmission across a series of $M$ potential steps. The incident  plane
wave with amplitude $A_0$ splits into a reflected and a transmitted component
with amplitudes $B_0$ and $A_{M+1}$, respectively.
}
\label{figDiracScheme}
\end{figure}

For not too narrow GNRs, the low energy excitations can be treated very
efficiently within the Dirac approximation~\cite{Wallace1947,CastroNeto2009}.
Boundary conditions of GNRs require the wave function to vanish on the
(fictitious) sites just outside the GNR, \ie, at $y=0$ and $y=W+a_0$, where the
$y$ axis is perpendicular to the direction of the GNR and the lower edge of the
GNR is located at $y=a_0/2$ [see Fig.~\ref{figDiracScheme}({a})], where $a_0$ is
the lattice period %distance between neighboring C atoms 
along the $y$ direction. In the case of
armchair GNRs, this affects both sublattices and the boundary conditions can be
fulfilled by a superposition of two states from different valleys with the same
energy  $E=\hbar\vF (k_\perp^2 + \smash{k_\parallel^2})^{1/2}$ and equal
longitudinal momentum $\hbar \k_\parallel$, but with opposite transverse
momentum $\pm \hbar \k_\perp$, measured from the Dirac points
\cite{Brey2006,Wakabayashi2009}.   Here $\vF$ is the Fermi velocity in graphene.
Note that the  effective description given by the Dirac equation holds as long
as the  $\vc{k}\cdot \vc{p}$ approximation remains valid, \ie, for not too
narrow GNRs.

Since the valley momenta $\K$ and $\K'$ can be chosen parallel to $\k_\perp$,
the transverse wave function can be written   $\phi_\perp(y)=
\sin\bigl[(K+k_\perp) y \bigr]$ where $K=4\pi/3a_0$. This function is
evaluated on the honeycomb lattice with   $y \in \mathbb{N} a_0/2$ and
oscillates rapidly. The transverse momentum $\k_\perp$,  however, is small and 
quantized by the conditions $\phi_\perp(W + a_0) = \phi_\perp(0)= 0$. The
allowed values for $k_\perp$ are given by $(K+k_{\perp n})(W+a_0)=\mathbb{Z} \pi$, and the spectrum reads
\begin{equation}
 E_n(k_\parallel) = \pm\hbar \vF \sqrt{k_{\perp n}^2 + \smash{k_\parallel^2}}\ .
\end{equation} 

Taking into account that $W$ is an integer multiple of $a_0/2$, one finds that 
the spectrum is gapless if \cite{Wakabayashi2009}
\begin{equation}\label{gapless}
W = (3 n_1 +1)a_0/2\, , \quad n_1\in\mathbb{N}\ .
\end{equation}
For asymmetric armchair GNRs, as in Ref.\ \onlinecite{Brey2006}, $n_1$ is even. 
For symmetric armchair GNRs, $W$ is an integer multiple of $a_0$ and $n_1$ is
odd, such that $W = (3 n - 1)a_0$, $n\in\mathbb{N}$ implies a gapless spectrum.

In real samples there are small gaps even in the case~\eqref{gapless}, which are
due to edge effects \cite{Han2007,Son2006} not included in the simple Dirac
ansatz nor the homogeneous tight-binding formulation.  In this work we consider
symmetric armchair GNRs of width $W=\widetilde n a_0$, where the integer
$\widetilde n$ is different from $3 \mathbb{N} - 1$, e.g., $W=40a_0$.  
In this case, there is a band gap already due to
the above reasoning, and the edge effects are negligible.
Then, the allowed
values of the transverse momentum are
\begin{align}
 |k_{\perp n}| &= \frac{\pi n}{3(W+a_0)}\ ,\quad n = 1,2,4,5,7,8,\,\ldots\, ,
\end{align}
and the half-gap is $E_0=E_1(0)= \pi \hbar\vF / [3(W+a_0)] = \unit[61.9]{meV}$.
In the following, we will consider only the lowest transverse momentum $k_{\perp1}$ and omit the index 1.

\subsection{Transfer-matrix description of transmission}\label{appTransfer}

For potentials depending only on the longitudinal coordinate $x$,
the  transverse momentum $\k_\perp$ together with the wave function
$\phi_{\perp}(y)$ is conserved, and it suffices to solve for the longitudinal
wave function $\phi_\parallel(x)$. We  consider the transmission across a
piecewise constant potential profile, as sketched in
Fig.~\ref{figDiracScheme}(b).  The solution of the Dirac equation for each spin
$\sigma=\pm 1$ and in each interval of constant potential value $V$ is the
superposition of two counter-propagating sublattice pseudo spinors
\begin{equation}
  \psi_{\parallel}(x) = 
  A \begin{pmatrix}
  e^{-i \theta/2} \\
 -e^{ i \theta/2}
 \end{pmatrix}
 e^{i k_\parallel x}
+ B  \begin{pmatrix}
  e^{+i \theta/2} \\
 -e^{- i \theta/2}
 \end{pmatrix}
 e^{-i k_\parallel x} \ ,
\label{eqWavefunction}
\end{equation}
with $\tan \theta = k_\parallel/k_\perp$ and $k_\parallel = [(E-V)^2/(\hbar
\vF)^2 - k_\perp^2]^{1/2}$. The solution may be evanescent  because
Eq.~\eqref{eqWavefunction} holds also for $|E-V|<\hbar\vF |k_\perp|$,  when
$k_\parallel$ and $\theta$ become imaginary. Then, the general form of the wave
function in each slab $j$ with potential $V_j$ and momentum $k_\parallel = k_j$
is 
\begin{align}
 \begin{pmatrix}
  e^{-i \theta_j/2} &  e^{ i \theta_j/2}\\
 -e^{ i \theta_j/2} & -e^{-i \theta_j/2}
 \end{pmatrix}
 \begin{pmatrix}
  A_j(x) \\
  B_j(x)
 \end{pmatrix} =: S_j
 \begin{pmatrix}
  A_j(x) \\
  B_j(x)
 \end{pmatrix}\ ,
\end{align}
where $A_j(x)= A_j e^{i k_j x}$ and $B_j(x) =B_j e^{-i k_j x }$, such that
\begin{align}\label{eqFreePropagation}
 \begin{pmatrix}
  A_j(x_{j+1}) \\
  B_j(x_{j+1})
 \end{pmatrix} 
 = {G_j}
\begin{pmatrix}
A_j(x_{j}) \\
 B_j(x_{j})
\end{pmatrix} ,
\end{align}
with ${G_j}=e^{i k_j (x_{j+1}-x_j) \sigma_z}$.  
At each junction, $k_j$ changes but the wave function
remains continuous: 
\begin{align}\label{eqPotentialStep}
S_j
\begin{pmatrix}
  A_j(x_j) \\
  B_j(x_j)
 \end{pmatrix}
=
S_{j-1}
\begin{pmatrix}
  A_{j-1}(x_j) \\
  B_{j-1}(x_j)
 \end{pmatrix} .
\end{align}
With the help of Eqs.~\eqref{eqFreePropagation} and~\eqref{eqPotentialStep},
one writes down the transfer matrix for the whole system
\begin{align}\label{eqTransferMatrix}
\begin{pmatrix}
  A_{M+1} \\
  B_{M+1}
 \end{pmatrix}
 = S_{M+1}^{-1} \widetilde G_{M} \ldots  
 \widetilde G_{2} \widetilde G_{1} S_0 
\begin{pmatrix}
  A_{0} \\
  B_{0}
 \end{pmatrix}\ ,
\end{align}
with $\widetilde G_{j} = S_j G_j S_j^{-1}$. 

For the transmission problem depicted in Fig.~\ref{figDiracScheme}({b}),  the
boundary condition is no incoming electron from the right, $B_{M+1}=0$. The
reflection probability at the left is the ratio of reflected to incident
current, $R=|B_0|^2/|A_0|^2$.  For the transmission probability one has to take
into account that the longitudinal momenta $k_{M+1}$ and $k_0$ are different if
$V_0 \neq V_{M+1}$, such that the ratio of transmitted to incident current is
$T=(|A_{M+1}|^2 k_{M+1})/(|A_0|^2 k_0)$.

\subsection{Band structure of the unbiased lattice}\label{appBandstructure}

The Dirac formalism allows us to analytically study the system in the limit
$N\rightarrow\infty$, when the energy regions with high transmission become
transmission bands surrounded by insulating bands %energy regions 
with $T=0$.  For an unbiased
lattice with identical barriers of width $a$ and spacing $b$, there are only two
different transfer matrices involved, $\widetilde G_a$ and $\widetilde G_b$. In
the limit $N\to\infty$, the superlattice  eigenfunctions have the Bloch phases
$\exp(\pm i q l)$, that are the eigenvalues of the transfer matrix $\widetilde
G= \widetilde G_a \widetilde G_b$ over one lattice period $l=a+b$.  Thus, the
dispersion relation $E(q,k_\perp)$ is obtained as $\cos(q l) = \Tr(\widetilde
G)/2$, or again~\cite{Barbier2010,Zhao2012}
\begin{eqnarray}
\label{eqSuperlatticeBands}
\cos q l &=& \cos k_a a  \cos k_b b \nonumber \\ 
&+& \frac{\cos\theta_a\cos\theta_b -1}{\sin\theta_a\sin\theta_b}\,
\sin k_a a\, \sin k_b b \ . 
\end{eqnarray} 
If $|\Tr(\widetilde G)/2| >1$, then there is no propagating solution with
real-valued $q$, and $E$ falls into the bandgap of the superlattice. In Fig.~\ref{figZeroBias}(a), the transmission bands for both spin channels are indicated by the extended bars on the bottom. 

\subsection{Spin-polarized current at finite bias}

Because the superlattice potential depends on the carrier spin, the transmission
probability $T_\pm$ is also spin-dependent.  Hereafter, $+$ ($-$) signs and red
(blue) colors in all figures correspond to spin-up (spin-down) electrons
respectively.  In order to  to calculate
the spin-dependent electric currents $I_\pm$ across the 
sample from the transmission probabilities $T_\pm$ , we use the Landauer-B\"{u}ttiker scattering formalism \cite{Buttiker1985}
\begin{equation*}
 I_\pm=\frac{2e}{h} \int 
T_\pm(E,V_{\rm SD})\,\Big[f(E-\mu_\mathrm{S})-f(E-\mu_\mathrm{D})\Big] 
\mathrm{d}E\ , 
\end{equation*}
where $f(\epsilon) = [ \exp(\epsilon/\kB T) + 1 ]^{-1}$ is the Fermi-Dirac
distribution at temperature $T$. We address the current and its polarization at
a finite bias voltage $V_\mathrm{SD}$ between source and drain, whose chemical
potentials, $\mu_\mathrm{S} = \mu + e V_\mathrm{SD}$ and $\mu_\mathrm{D}=\mu$,
have the same offset $\mu$ from the Dirac point.
Using $I_{+}$ and $I_{-}$ we can calculate the total current $I=I_{+}+I_{-}$ through the
device, as well as its spin-polarization $P=(I_{+}-I_{-})/I$. 

\section{Results}\label{secResults}

\subsection{Transmission at zero bias}

\begin{figure}[tb]
\includegraphics[width=0.90\columnwidth,clip]
{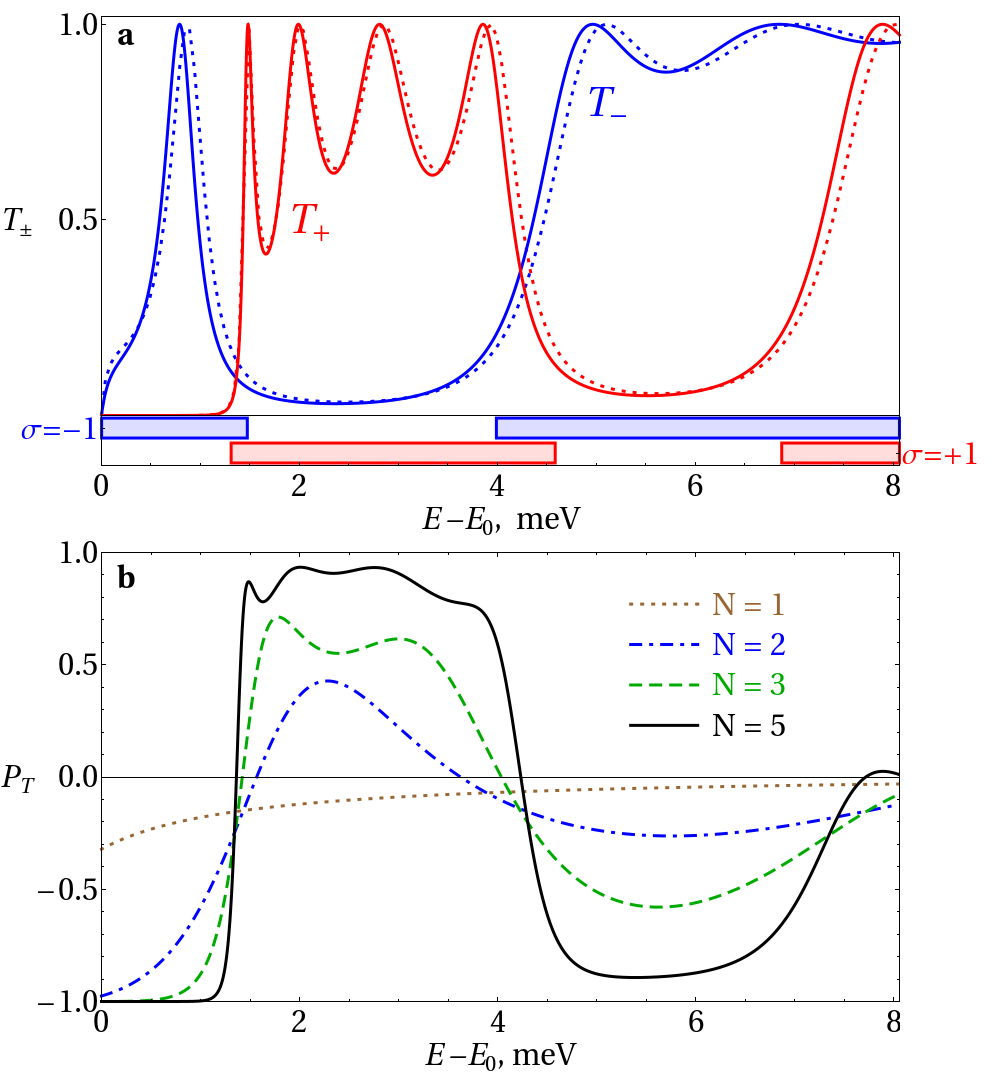}
\caption{(Color online)
(a)~Transmission probabilities $T_\pm$ as functions of energy for spin-up (red)
and spin-down (blue) electrons.  There is a very good agreement between the
Dirac approximation (solid lines) and the tight-binding calculation (dotted
lines). Horizontal bars in the lower part of the figure indicate the energy
bands of the infinite superlattice, obtained from~\eqref{eqSuperlatticeBands}. 
(b)~Transmission polarization $P_T$ as a function of energy for different
numbers $N$ of ferromagnetic strips.
}
\label{figZeroBias}
\end{figure}

Figure~\ref{figZeroBias}(a) shows the transmission probability
through the unbiased sample calculated within the Dirac approximation (solid
lines) and the full tight-binding model (dotted lines). % which is more accurate.
Already for the relatively small number $N=5$ of strips, regions of high transmission 
coincide quite well with the bands of the infinite superlattice introduced in Sec.~\ref{appBandstructure} (and indicated by the horizontal bars at the bottom of the figure). The origin of the energy for each curve is set to the lowest subband bottom
energy $E_0$ calculated within the corresponding model.  The figure demonstrates
very good agreement between the two approaches.  The Dirac approximation
overestimates slightly the value of $E_0$ (by $0.3\%$) but %it 
is accurate enough
for our purposes.  Unless stated otherwise, in the following, we use the Dirac
approximation since it demands less computational resources.  

\begin{figure*}[tb]
 \includegraphics%[width=0.99\linewidth]
{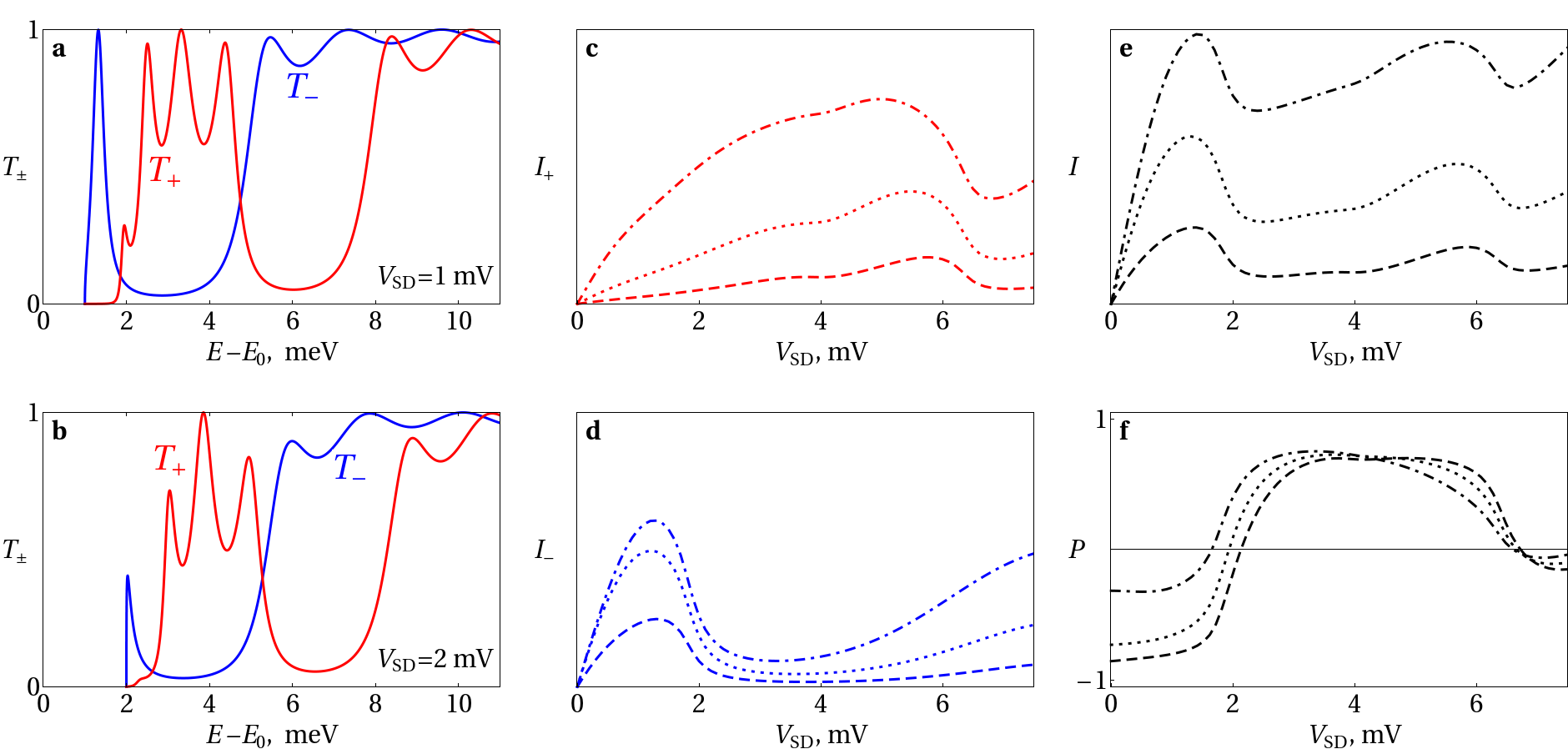}
\caption{(Color online)
Panels~(a) and~(b) show that the transmission bands for both spins at finite bias
$V_{\rm SD}$ are shifted, quenched and distorted compared to the unbiased case
[Fig.~\ref{figZeroBias}(a)]. Panels~(c) and~(d) display the spin-polarized currents
$I_\pm$ as functions of the bias $V_\mathrm{SD}$, for $T=\unit[4]{K}$ and
different values of the chemical potential: $\mu-E_0=\unit[0]{meV}$ (dashed),
$\unit[0.5]{meV}$ (dotted), and $\unit[1.0]{meV}$ (dash-dotted) lines. Finally,
panels~e) and~f) show the total current $I=I_{+}+I_{-}$ and the current
polarization $P=(I_{+}-I_{-})/I$ for the same parameters.  All the intensity
graphs use the same arbitrary scale.
}
\label{figFiniteBias}
\end{figure*}

The transmission is spin-dependent, which manifests itself clearly in the
transmission polarization, defined as $P_T =(T_{+}-T_{-})/(T_{+}+T_{-})$ and
shown in Fig.~\ref{figZeroBias}(b). As the number of strips is increased, the
transmission probability at energies outside the transmission bands vanishes
rapidly, thus leading to an enhanced polarization.  For $N \geq 3$, the
transmission polarization noticeably changes within narrow energy intervals.
Such abrupt polarization switching can be expected only if the overlap between
transmission bands corresponding to different spins is small, as seen in
Fig.~\ref{figZeroBias}(a). The overlap is determined by different factors: the
splitting $\Delta_\mathrm{ex}$ and the geometrical parameters $a$ and $b$, which
should be chosen carefully in order to observe a pronounced switching and
filtering effect in a real device.  Such a choice can be made, for example, by
analyzing the band structure of the infinite lattice within the Dirac
approximation given by~\eqref{eqSuperlatticeBands}.

\subsection{Spin-polarized current at finite bias}\label{secBias}

As depicted in the lower panel of Fig.\ \ref{figScheme}, we assume the bias
voltage to drop along the sample in a roughly Ohmic manner.  For simplicity,
we assume that the voltage drops occur at the edges of the EuO strips only,
resulting in a piecewise constant potential profile as shown in the lower
panel of Fig.~\ref{figScheme}.  
Such a model potential allows us to use the
efficient Dirac transfer matrix method discussed in Sec.~\ref{appTransfer}. 
There may be additional voltage drops at the source and drain contacts, which are just
outside the middle and lower panels of Fig.~\ref{figScheme}.  With the term
\emph{bias voltage} $V_{\rm SD}$, we refer only to the voltage drop across
the GNR.  The exact potential profile can in principle be obtained from a
self-consistent electrostatic potential calculation, but that would go
beyond the scope of this work.  We note that, since the desired potential
profile $\epsilon_i$ is spin independent, it can always be adjusted via gate
voltages.
In Appendix \ref{appDifferentLadders} we demonstrate that the spin-dependent transmission does not depend crucially on the details of the biased potential profile.

The bias results in a distortion of the transmission bands: the bands shift, quench and finally disappear as the voltage increases, as
is seen in Fig.~\ref{figFiniteBias}(a) and (b) compared to Fig.~\ref{figZeroBias}(a). The polarized currents $I_\pm$ as
functions of $V_{\rm SD}$ are plotted in Fig.~\ref{figFiniteBias}(c) and~(d). 
The spin-dependent transmission bands and their distortion due to the bias lead
to NDR regions at different values of the bias voltage for different spins.  For
spin down, the NDR occurs at a lower bias and the negative slope of the
current-voltage curve is particularly steep, which is due to the fact that the
first transmission peak remains very sharp until it disappears
[Fig.~\ref{figFiniteBias}(a) and~(b)].  The lowest spin-up transmission band
gets washed out before it disappears at a higher bias, resulting in the less
pronounced NDR.

We further address the total current $I$ through the device, as well as its
spin-polarization $P$.  Figures~\ref{figFiniteBias}(e)~and~(f) show that the
total current $I$ also manifests NDR for two different biases, corresponding
to the NDR regions of $I_-$ and $I_+$.  The current polarization shows an
initial range with negative values followed by a second region dominated by
the spin-up current.  As the bias increases further, the polarization decays
and finally vanishes.  Note that the current is highly polarized for certain
biases, which proves that the device can operate as a spin filter.  On the
other hand, because the characteristics $I_+(V_\mathrm{SD})$ and
$I_-(V_\mathrm{SD})$ are very different, if the source feeds partially
polarized electrons, the total current through the device would depend on
the degree of the electron polarization.  The latter opens a possibility to
determine the polarization of a current by purely electrical measurements,
which is a very promising application.

We have considered an ideal device with perfect rectangular GNR and strips,
while different imperfections and perturbations can introduce disorder into the
system and affect the electric current and its polarization~\cite{Zhao2012}. 
There are different possible sources of disorder, for example, charged
impurities in the substrate or defects of the device fabrication.  The former
results in an additional smooth electrostatic potential and can hardly
deteriorate the transmission through the device to a large extent. However, the
effect of the latter on the transport properties can be stronger. To estimate
the possible impact of the fabrication imperfections on the predicted effects,
we considered disordered superlattices with randomly varying strip widths and
spacings, up to 20\%. Our calculations (not shown here) demonstrated that the
transmission bands are affected by the disorder to a comparable degree for both
spin up and spin down electrons, which suggests that a moderate disorder would
not seriously deteriorate transport and polarization properties of the device. 
The current magnitude remains almost the same, and the NDR turns out to be
robust under the effects of disorder as well.

\section{Conclusions} \label{conclusions}

In summary, we propose a novel graphene-based device comprising a GNR and a
regular array of ferromagnetic strips on top of it.  The ferromagnets induce a
proximity exchange splitting of the electronic states in the GNR and create a
spin-dependent superlattice.  We have shown that the electric current through
the device can be highly polarized.  Thus, the device can operate as a spin
filter.  Alternatively, it can be used to obtain the polarization degree of the
source electrons by purely
electrical measurements.  Moreover, the two polarized components of the current
manifest non-monotonic dependencies on the bias voltage.  In particular, for
both spins, the current-voltage characteristics present regions with negative
differential resistance for the bias in the range of a few millivolts.  The
device operates therefore as a low-voltage Esaki diode for spin-polarized
currents.

An important advantage of the superlattice induced by ferromagnets is that the
exchange interaction is very short-ranged; its characteristic length scale is on
the order of one monolayer. Unlike the long-range electrostatic gate potentials
which can interfere with each other, setting a practical lower limit for the
inter-device spacing, the exchange-interaction induced potential profiles are
very abrupt.  Therefore, heterostructures created by ferromagnets allow for very
close packing of circuits and, consequently, considerably higher device
densities.

Finally, we note that in a spintronic device the degree of freedom that carries
information is the polarization of the current rather than its magnitude.  We
have shown that the current polarization is also a non-monotonic function of the
bias voltage, suggesting that the superlattice can be used as a Esaki spin 
diode.  This opens a possibility to design a whole new class of spintronic
circuits such as spin oscillators, amplifiers and triggers.

\vspace{2em}

\begin{acknowledgments}

Work in Madrid was supported by the MICINN (project MAT2010-17180). Research of
C.G.\ was supported by the PICATA postdoctoral fellowship from the Moncloa
Campus of International Excellence (UCM-UPM). P.~A.~O acknowledges financial
support from the FONDECYT (grant 1100560). CQT is a Research Centre of
Excellence funded by the Ministry of Education and the National Research
Foundation of Singapore.

\end{acknowledgments}

\appendix

\section{Robustness against details of the potential profile}\label{appDifferentLadders}

\begin{figure}[h]
 \includegraphics%[width=0.99\linewidth]
{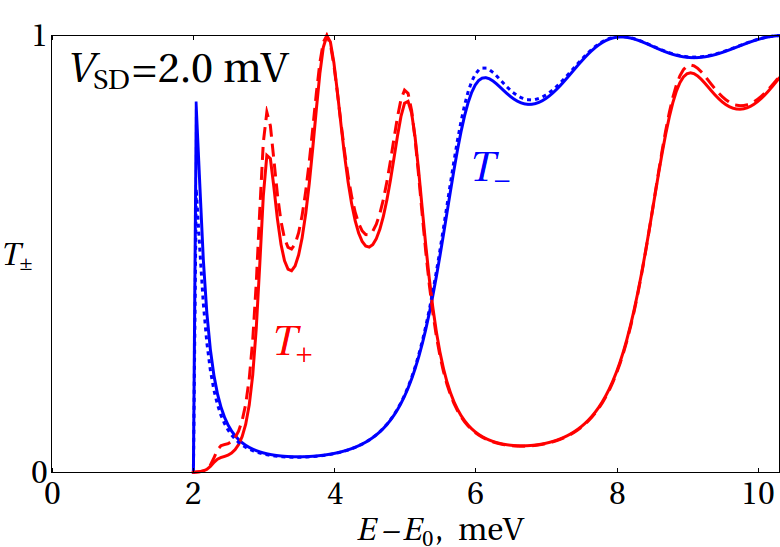}
\caption{(Color online)
Transmissions $T_\pm$ for both spins at finite bias $V_\mathrm{SD} = \unit[2]{mV}$, assuming
that the voltage drop occurs at the edges of the EuO strips (solid) or linearly 
along the sample (dashed).  Here, we use the tight-binding model, which can be 
compared with the Dirac approximation on Fig.~\ref{figFiniteBias}(b).
}
\label{figDifferentLadders}
\end{figure}

To estimate the accuracy of the transmissions $T_\pm$ obtained using the
piecewise constant potential of the main text, we compared them with those calculated using a different potential profile, where a linear gradient $V_{\rm SD}/(5a+6b)$ is added to the spin-dependent superlattice potential of the middle panel of Fig.\ \ref{figScheme}.
Results obtained with the tight-binding method are shown in Fig.\ \ref{figDifferentLadders}.
There are only slight deviations, which proves that the details of the potential are not important for our findings and that the piecewise constant potential is a very good approximation. 

\bibliography{referencesNDR}%,supplement}

\end{document}